# A protein network refinement method based on module discovery and biological information

Li Pan, Haoyue Wang*, Jing Sun, Bin Li, Bo Yang and Wenbin Li*

**Abstract**—The identification of essential proteins can help in understanding the minimum requirements for cell survival and development. Network-based centrality approaches are commonly used to identify essential proteins from protein-protein interaction networks (PINs). Unfortunately, these approaches are limited by the poor quality of the underlying PIN data. To overcome this problem, researchers have focused on the prediction of essential proteins by combining PINs with other biological data. In this paper, we proposed a network refinement method based on module discovery and biological information to obtain a higher quality PIN. First, to extract the maximal connected subgraph in the PIN and to divide it into different modules by using Fast-unfolding algorithm; then, to detect critical modules based on the homology information, subcellular localization information and topology information within each module, and to construct a more refined network (CM-PIN). To evaluate the effectiveness of the proposed method, we used 10 typical network-based centrality methods (LAC, DC, DMNC, NC, TP, LID, CC, BC, PR, LR) to compare the overall performance of the CM-PIN with those the refined dynamic protein network (RD-PIN). The experimental results showed that the CM-PIN was optimal in terms of precision-recall curve, jackknife curve and other criteria, and can help to identify essential proteins more accurately.

——————————— ◆ ———————————

## 1 INTRODUCTION

PROTEINS are the most significant components of living organisms and have very important biological functions, participating in gene regulation, cellular metabolism, and are the main bearers of biological life activities. Proteins are subdivided into essential and non-essential proteins, among which, essential proteins are particularly important for life activities, and their absence can lead to the failure of the organism to survive [1]. In addition, essential proteins are associated with human disease-causing genes, and their identification and analysis can help the protein design of drug targets. However, early studies of essential proteins were mainly conducted by wet experimental methods such as RNA interference [2], single gene knockout [3] and conditional gene knockout [4], which often have the drawbacks of being expensive and time-consuming, therefore, the identification of essential proteins by computational methods has become the current trend.

Network topology-based centrality methods are commonly computed to identify essential proteins [5], such as: degree centrality (DC) [6], local average connectivity centrality (LAC) [7], node clustering centrality (NC) [8], maximum neighborhood component density centrality (DMNC) [9], topological potential centrality (TP) [10], neighbor interaction density centrality (LID) [11], closeness centrality (CC) [12], betweenness centrality (BC) [13], pagerank centrality (PR) [14], leaderrank centrality (LR) [15], etc. Although these network-based centrality methods have made great progress in the identification of essential proteins, they are highly dependent on the accuracy of protein interaction networks (PINs). However, a large proportion of PINs obtained from high-throughput biological experiments have been found to contain false positives [16], which can affect the quality of PINs and thus interfere with the recognition rate of essential proteins by centrality methods.

To address the impact of false positives on the recognition rate of essential proteins, some researchers have improved the recognition rate of essential proteins by fusing biological information to eliminate the noise in PINs. For example, Xiao et al [17] combined PINs with gene expression level data by observing whether protein pairs were activated at the same time, taking into account the temporal factor in PINs, and proposed a once refined PIN (dynamic protein interaction network), which improved the identification accuracy of the centrality method for essential proteins. Subsequently, based on dynamic protein interaction network, Li et al [18] added spatial characteristics in PINs by observing whether protein pairs appeared in the same subcellular compartment and constructed a secondary refined PIN (refined dynamic protein interaction network) by using gene expression level data and subcellular localization information, and the centrality methods were able to identify more essential proteins on the secondary refined network compared with the once refined dynamic network. However, some researchers pointed out that PINs have modular characteristics [19], [20], and proteins within modules have

————————————————
• L. Pan, H. Y. Wang, J. Sun, B. Li, B. Yang and W. B. Li are with Hunan Institute of Science and Technology, Yueyang, Hunan, China. E-mail: {haoyue_wang111@163.com, wenbin_lii@163.com}.
• *Corresponding author

higher similarity than those in different modules. The construction of the aforementioned networks focused only on the edges of the protein nodes in the network, ignoring the modularity feature of PINs. Hart et al [21] pointed out that the essentiality of a protein is not only related to the protein itself, but also to the functional module in which the protein is located. Furthermore, Zotenko et al [22] found that in PINs, a large number of essential proteins may be present in highly dense functional modules. Therefore, some other researchers have started to propose new methods for essential protein identification by combining different biological information and modular features of PINs. For example, Qin et al [23] proposed LBCC, an essential protein metric method based on network topological features and protein complexes (functional modules); Li et al [24] pointed out that proteins in protein complexes are more likely to be essential than any proteins not present in the complexes, and they proposed a centrality method UC by combining protein complexes and topological features of PINs; Lei et al [25] proposed an essential protein identification method PCSD that fuses the degree of protein complex involvement and subgraph density; Lu et al [26] proposed two new methods, CDC and CIBD, by combining local features of protein complexes and topological properties of PINs to determine protein importance. Although these methods can identify more essential proteins, they still rely on the accuracy of PINs, and most of them will use known protein complexes to identify essential proteins, with non-essential proteins in protein complexes sometimes being misidentified as essential. Therefore, it is still a question worth exploring how to better utilize the modularity feature of PINs to identify essential proteins.

However, in order to discover the functional modules in different PINs, some researchers detected the protein functional modules by clustering proteins, and pointed out that this is an effective way to discover functional modules of proteins [19], [27]. Clustering aims to group proteins into clusters so that proteins are more likely to interact within clusters than outside of them. In our study, we found that the biological and topological information in different modules divided by hierarchical clustering methods is diverse and there are some dense modules that still contain a lot of noise, and even some modules which may be non-critical that do not contain essential proteins. It can be seen that it is feasible to refine the network by combining the modular characteristics of PINs to improve the identification accuracy of essential proteins, and it is important to determine and select the critical modules for constructing higher quality PINs.

Based on this, in this paper, we proposed a protein interaction network refinement method based on module discovery and biological information for improving the identification accuracy of essential proteins and its steps are as follow: firstly, gene expression level data and subcellular localization information were used to construct a secondary refined protein network, secondly, remove the interactions in some small connected subgraphs from the secondary refined protein network, and divide the maximal connected subgraph into several closely connected modules by the Fast-unfolding algorithm [28] that fuses the modularity; thirdly, select the critical modules by combining protein homology score, subcellular localization information and topological features of each module; fourthly, if there are connected edges between the proteins in the original network and they are in the critical modules, retain them, otherwise remove them from the edge set, and the obtained PIN is represented by CM-PIN. To demonstrate the effectiveness of this method, three networks were tested in this paper, namely, the static protein interaction network (S-PIN) [29], the dynamic protein interaction network (D-PIN) [17] and the refined dynamic protein interaction network (RD-PIN) [18]. We compared the identification accuracies of essential proteins in the above three network and CM-PIN by using 10 centrality methods (LAC, DC, DMNC, NC, TP, LID, CC, BC, PR, LR), and also plotted the precision-recall curves for each centrality method and calculated the area under the curves. To further validate the overall effectiveness of the CM-PIN, the jackknifing method was used for comparison and six evaluation metrics were calculated for each centrality method: sensitivity, specificity, positive predictive value, negative predictive value, F-measure, and accuracy. The results showed that the 10 centrality methods for the recognition rate of essential proteins and their performances are the best in CM-PIN. Finally, we used another dataset to verify the effectiveness of our method and verified whether the network refinement method proposed in this paper could be effectively applied to networks containing more noise, the results also demonstrated that the CM-PIN could help identify more essential proteins, and our method could help to filter false positives in PINs whether they are networks with a lot of noise or refined networks.

## 2 METHODS

In this section, first, we described how to build these three protein interaction networks: S-PIN, D-PIN, and RD-PIN. Second, we described how to screen the critical modules by the biological information of proteins and the topological features of each module, and construct a more refined protein interaction network based on RD-PIN, and the overall steps of the method were shown in Figure 1.

### 2.1 Construction of S-PIN, D-PIN and RD-PIN

(1) Static protein interaction network (S-PIN) [29], [30]

Based on the yeast database of interaction proteins (DIP dataset), we construction an undirected graph $G_S = (V_S, E_S)$, where $V_S$ represents the set of proteins and $E_S$ represents the set of protein interactions.

(2) Dynamic protein interaction network (D-PIN) [17], [31]

Based on the gene expression level data, which is normally distributed and has been normalized, first, the activity of each protein was calculated at 36 time points. The activity threshold $\tau_i$ of protein $v_i$ was calculated by using the following equation [19], [32]:

$$\tau_i = \mu_i + \sigma_i \quad (1)$$

where $\mu_i$ denotes the mean of the 36 time-point gene ex-pression level values of the protein and $\sigma_i$ is the standard deviation of the gene expression level values of $v_i$. Then, let $e_{ik}$ denotes the value of gene expression level of $v_i$ at time point $t_k$. If $e_{ik}$ is greater than $\tau_i$, then $v_i$ is active at time point $t_k$. $\forall (v_i, v_j) \in E_S$, if both $v_i$ and $v_j$ are activated at time point $t_k$, the interaction between them is preserved, otherwise it is removed from $E_S$, thus obtain the refined $E_D$ and construct the undirected graph $G_D = (V_D, E_D)$, where $V_D = V_S$.

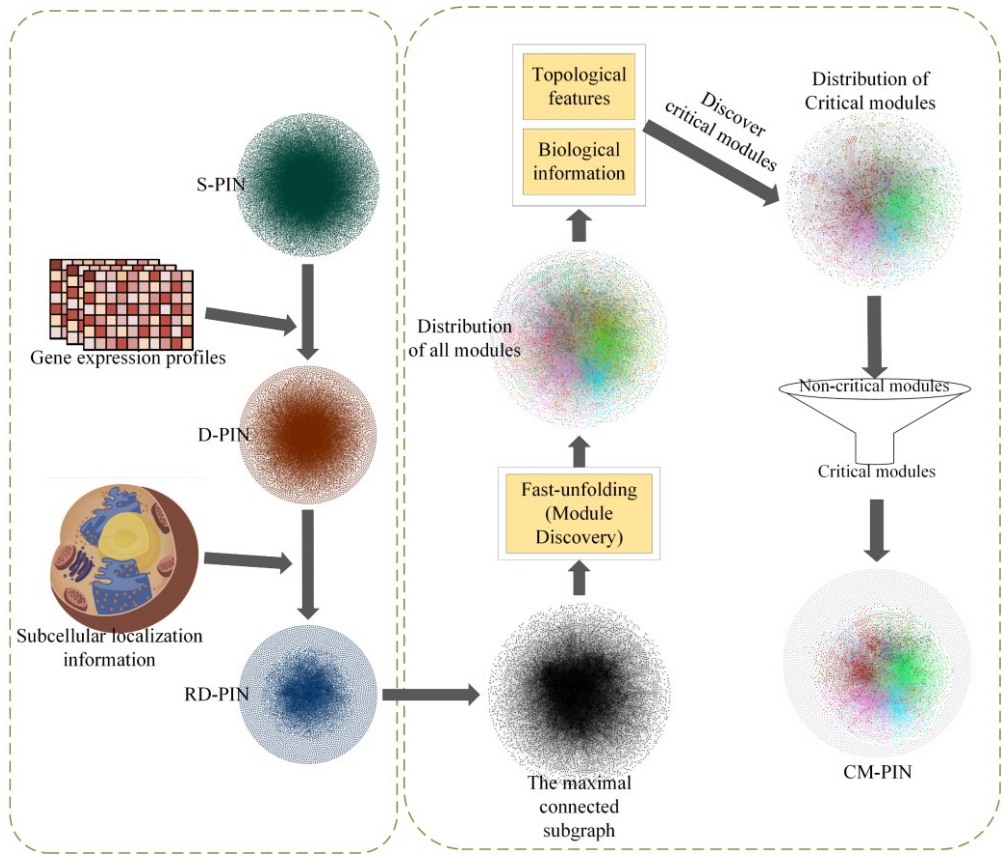

Fig.1 The overall steps of the CM-PIN construction method.

(3) Refined dynamic protein interaction network (RD-PIN) [18]

Based on the 11 subcellular localization information of the proteins, for protein $v_i$, let $L(v_i) = \{l_1, ..., l_m, ..., l_r\}$, where $r = 11$, if $v_i$ is in the $m$th subcellular compartment, $v_i(l_m) = 1$. $\forall (v_i, v_j) \in E_D$, only when $v_i(l_m) = v_j(l_m) = 1$, the $v_i$ and $v_j$ will interact, otherwise their interaction will be removed from the $E_D$, thus obtain the twice refined $E_{RD}$, and construct the undirected graph $G_{RD} = (V_{RD}, E_{RD})$, where $V_{RD} = V_S$.

## 2.2 Construction of CM-PIN

### 2.2.1 Retaining interactions in maximal connected subgraphs

It has been found that PINs have scale-free properties [33], [34]. The scale-free property means that the degrees of the nodes in PIN obey a power-law distribution, so PIN belongs to a scale-free network. Considering that PIN is a disconnected graph and consists of several connected subgraphs, where most of the proteins and their interactions are present in a maximal connected subgraph, while the number of proteins and their interactions in some remaining connected subgraphs are very small. As shown in Table 1, we counted the proportion of interactions in the maximal connected subgraphs of the DIP and BioGRID datasets to the original network interactions. Therefore, in order to obtain more accurate module partitioning in downstream work, based on the robustness of the scale-free network, we retained the interactions in the largest connected subgraph of RD-PIN, in other words, we removed the interactions in the remaining smaller connected subgraphs of RD-PIN.

Table 1. The proportion of interactions in the maximal connected subgraphs to the original network interactions on DIP and BioGRID datasets.

| Networks | DIP | BioGRID |
|---|---|---|
| S-PIN | 15123/15166=99.72% | 52832/52833=99.99% |
| D-PIN | 9436/9514=99.18% | 32730/32735=99.98% |
| RD-PIN | 4953/5175=95.71% | 18330/18362=99.83% |

### 2.2.2 Module discovery based on Fast-unfolding algorithm

It has been shown that PINs have modular properties [19], [20], and modularity reflects the presence of highly connected protein clusters in PINs. So far, clustering of protein interaction networks is an effective method for module delineation, and in this paper, Fast-unfolding module discovery algorithm, a hierarchical clustering method, is used for module delineation of RD-PIN.

The purpose of module partitioning is to make the connections within the partitioned modules tighter and the connections between modules sparser. In order to evaluate whether the module division is feasible, Newman et al [35] proposed the concept of modularity. Defining $e_{ii}$ as the ratio of the sum of all connected edges within module $i$ to the total number of edges in the network and $a_i$ as the ratio of the total number of neighboring nodes of nodes within module $i$ to the total number of edges, the modularity $Q$ can be expressed as:

$$Q = \sum_i (e_{ii} - a_i^2) \quad (2)$$

A larger modularity represents a tighter connection within the module, and conversely, a smaller modularity represents a sparser connection within the module, and when the modularity $Q$ reaches its maximum value, the division of modules is optimal.

Blondel et al [36] proposed a Fast-unfolding algorithm for discovering module structures on large networks, which is a heuristic algorithm based on modularity optimization. Compared with traditional module discovery algorithms, Fast-unfolding has lower time complexity on large-scale networks and stable results for module partitioning, which is the reason why this algorithm is chosen to partition modules in this paper. The implementation steps of Fast-unfolding algorithm are as follows: first, initialization, divide each protein node into different modules; second, for each protein node, try to divide it into the module where its neighboring nodes are located, calculate the modularity $Q$ at this time, and judge whether the difference $\Delta Q$ between the modularity before and after the division is positive, if it is positive, accept this division, if not, abandon this division; third, repeat the above process until the modularity $Q$ can no longer be increased, then the division of modules is completed, and $C = \{c_1, c_2, ..., c_i, ..., c_m\}$ is the set of modules and $m$ is the number of module divisions. It is worth noting that the divided modules are non-overlapping.

### 2.2.3 Detecting critical modules

There are differences between different modules in the protein interaction network, and some modules are crucial to the protein interaction network. To determine the criticality of each module, we used three features to score each module in the PIN: homology information, subcellular localization information, and topological information of the module in the PIN.

It was shown that the evolutionary rate of essential proteins is much slower than that of non-essential proteins [37], in other words, essential proteins are more conserved. To verify the conserved nature of essential proteins, researchers searched for homologous proteins from different species and calculated protein homology scores. In this paper, we used the protein homology scores to determine the criticality of modules, the idea is that first, each module is converted into a vector only contained 0 and 1 as a feature (each vector is a module, 1 if the protein is in that module, 0 otherwise), and calculate the Pearson correlation coefficient between each feature and the protein homology score, denoted as $Corr_i = \{corr_1, corr_2, ..., corr_m\}$, where $m$ is the number of modules. We think that the more correlated the module is with the protein homology score, the more likely it is to be critical. Thus, the set of possible critical modules selected based on the conservatism of the proteins within the module is denote as $C\_conservatism = \{c_i | Corr_i \geq th_1\}$, where $th_1$ is a threshold value.

The essentiality of proteins is not only related to protein homology scores, but also to the subcellular localization information of proteins. Proteins have different subcellular compartments, which play different roles and importance in cellular activities. To investigate the relationship between protein essentiality and subcellular localization information, we observed the number of occurrences of proteins and essential proteins in each subcellular compartment and found that proteins, even essential proteins, are most widely distributed in the nucleus, a subcellular compartment, so we concluded that the more the proteins in a module appear in the nucleus, the more likely the module is critical, denoted by $NSL$:

$$NSL(c_i) = \frac{N(c_i)}{n(c_i)} \quad (3)$$

where $N(c_i)$ is the number of times the protein within the module appears in the nucleus and $n(c_i)$ is the number of nodes within the module. The set of the possible critical modules selected based on the subcellular localization information of the proteins within the module is represented by $C\_subcellular = \{c_i | NSL(c_i) \geq th_2\}$, where $th_2$ is a threshold value.

To identify the criticality of the module, we also used the topological characteristics of each module in the network. It has been pointed out that a large number of essential proteins may exist in highly dense functional modules [22]. Thus, the richer the interactions within the module, the more likely it is to play an important role in the whole network, so we calculated the topological characteristics of each module $TF(c_i)$:

$$TF(c_i) = \frac{I(c_i) - O(c_i)}{n(c_i)} \quad (4)$$

where $I(c_i)$ is the number of interactions inside module $c_i$, $O(c_i)$ is the number of interactions between module $c_i$ and other modules, and $n(c_i)$ is the number of nodes of module $c_i$. And according to the topological characteristics of the module, the modules less than $th_3$ are selected as a non-critical module, in other words, the modules chosen here are potentially non-critical, i.e., $C\_topology = \{c_i | TF(c_i) \leq th_3\}$, where $th_3$ is a threshold value.

### 2.2.4 Building the MD-PIN

We integrated the above three features of the modules to obtain the final selected critical modules, i.e., $C\_critical = \{c_i | C\_conservatism \cup (C\_subcellular / C\_topology)\}$. $\forall (v_i, v_j) \in E_{RD}$, if $v_i$ and $v_j$ are both in the critical modules $C\_critical$, their interaction will be retained, otherwise their interactions will be removed from the $E_{RD}$, thus obtain the finally refined $E_{CM}$, resulting in a more refined protein interaction network CM-PIN, $G_{CM} = (V_{CM}, E_{CM})$, where $V_{CM} = V_{RD}$. Table 2 listed the topological characteristics of CM-PIN and other three networks in DIP dataset: the number of interactions, average degree, average clustering coefficient and density. It can be seen that CM-PIN is a more refined network (the construction process of CM-PIN is shown in the **Algorithm**).

Table 2. Topological characteristics of CM-PIN and other three networks on DIP dataset.

| Networks | The number of interactions | Average degree | Average clustering coefficient | Density |
|---|---|---|---|---|
| S-PIN | 15166 | 6.3911 | 0.0923 | 0.0013 |
| D-PIN | 9514 | 4.0093 | 0.0774 | 0.0008 |
| RD-PIN | 5175 | 2.1808 | 0.0848 | 0.0005 |
| CM-PIN | 3765 | 1.5866 | 0.0578 | 0.0003 |

## 3 EXPERIMENT AND DISCUSSION

### 3.1 DIP Dataset and bioinformatics datasets

All experiments in this paper are based on the Saccharomyces cerevisiae protein interaction dataset which is now the most complete data in all species and has been widely used to test various methods for the discovery of essential proteins. The yeast database of interaction proteins used in the experiment was downloaded from the database DIP [38], which contains 4,746 proteins and 15,166 interactions. The essential proteins are collected from the following datasets [39]: DEG, MIPS, SGD, SGDP, and the DIP dataset contains 1,130 essential proteins. Gene expression level dataset contains the gene expression level data of 36 time points of 6,777 proteins [40]. In addition, we downloaded protein homology information from the InParanoid database to calculate protein homology scores [41], [42], and downloaded subcellular localization data for proteins from the COMPARTMENTS database, which contains 11 subcellular locations: cytoskeleton, golgiapparatus, cytosol, endosome, mitochondrion, plasma membrane, nucleus, extracellular space, vacuole, endoplasmmic, reticulum, peroxisome [43].

### 3.2 Network-based centrality methods

The network-based centrality method will first calculate the centrality values of all protein nodes in the network according to its formula, then rank the proteins in descending order according to the centrality values, and finally the highly ranked proteins will be considered as essential proteins. To compare the S-PIN, D-PIN, RD-PIN with the CM-PIN, 10 typical network-based centrality methods were selected to identify essential proteins: DC, LAC, NC, DMNC, TP, LID, CC, BC, PR, LR. References for each centrality method and its formula are shown in Table 3.

### 3.3 Experimental results and analysis on DIP dataset

#### 3.3.1 Analysis of the number of essential proteins identification

In order to demonstrate that the network obtained by the method in this paper can effectively improve the number of essential proteins identified by each centrality method, we obtained a more refined and effective network (CM-PIN) and compared it with three networks (S-PIN, D-PIN, RD-PIN) on the number of essential proteins identification at top 100, top 200, top 300, top 400, top 500 and top

**Algorithm**: Construction of CM-PIN

**Input**: Static protein interaction network $G_S = (V_S, E_S)$, gene expression level data, subcellular localization information, protein homology information, thresholds for critical modules $th_1$, $th_2$, $th_3$

**Output**: A more refined and effective network $G_{CM} = (V_{CM}, E_{CM})$

**Begin:**
1: Construction of D-PIN $G_D = (V_D, E_D)$;
2: Construction of RD-PIN $G_{RD} = (V_{RD}, E_{RD})$;
3: $c_1, …, c_i, …, c_m \leftarrow$ Hierarchical clustering ($G_{RD}$);
4: **for** $i$ = 1 to $m$ **do**:
5:   $Corr(c_i) \leftarrow$ Pearson correlation coefficient ($c_i$, protein homology score);
6:   $NSL(c_i) \leftarrow$ the average number of times the protein in $c_i$ appears in the nucleus;
7:   $TF(c_i) \leftarrow$ the closeness of protein interactions in $c_i$;
8: **end for**
9: **for** $i$ = 1 to $m$ **do**:
10:  if $Corr(c_i) \geq th_1$:
11:    $C\_conservatism \leftarrow c_i$;
12:  end if
13:  if $NSL(c_i) \geq th_2$:
14:    $C\_subcellular \leftarrow c_i$;
15:  end if
16:  if $TF(c_i) \leq th_3$:
17:    $C\_topology \leftarrow c_i$;
18:  end if
19: **end for**
20: $C\_critical \leftarrow C\_conservatism \cup (C\_subcellular / C\_topology)$;
21: **for** each edge $(u, v) \in E_{RD}$ **do**:
22:  if edge $(u, v) \notin C\_critical$:
23:    remove edge $(u, v)$ from $E_{RD}$;
24:  end if
25: **end for**
26: **return** edge set $E_{CM} \leftarrow E_{RD}$ and $V_{CM} \leftarrow V_{RD}$;
27: **return** CM-PIN $G_{CM} = (V_{CM}, E_{CM})$.

Table 3. 10 network-based centrality methods and references to their formulas.

| | Centrality methods | Abbreviation | References to formulas |
|---|---|---|---|
| 1 | degree centrality | DC | [6] |
| 2 | local average connectivity centrality | LAC | [7] |
| 3 | node cluster centrality | NC | [8] |
| 4 | density of maximum neighborhood component | DMNC | [9] |
| 5 | topological potential centrality | TP | [10] |
| 6 | local interaction density | LID | [11] |
| 7 | closeness centrality | CC | [12] |
| 8 | betweenness centrality | BC | [13] |
| 9 | pagerank centrality | PR | [14] |
| 10 | leaderrank centrality | LR | [15] |

600, as shown in Figure 2. It can be seen that CM-PIN can significantly improve the identification accuracy of essential proteins and the curves from top 100 to top 600 are consistently above the other three networks. Compared with the best RD-PIN, the centrality methods in CM-PIN all have more than 6.2% improvement at top 600, and even some centrality methods have a significant im-provement, for example, the DMNC method has improved by 14.2% at top 600, and the LR method has improved by 15.1% at top 600. More importantly, the LID method, with 405 essential proteins identified at top 600, and as far as we know, there are few methods that can make the identification accuracy of essential proteins of DIP exceed 400 at top 600, which indicated that the network refinement method in this paper can effectively improve the identification accuracy of essential proteins. In addition, our method can improve the identification accuracy of essential proteins at top 600 of DIP by 12.7% to 40.1% for S-PIN, 8.0% to 38.3% for D-PIN and 6.2% to 15.1% for RD-PIN, all of which illustrated the effectiveness of our method and demonstrate that CM-PIN is a more refined and effective network.

significant, whether it is neighborhood-based, path-based or eigenvector-based centrality methods. This further demonstrates that the network refinement method in this paper is effective in removing noise and false positives from protein interaction networks and proves that CM-PIN is a more efficient network.

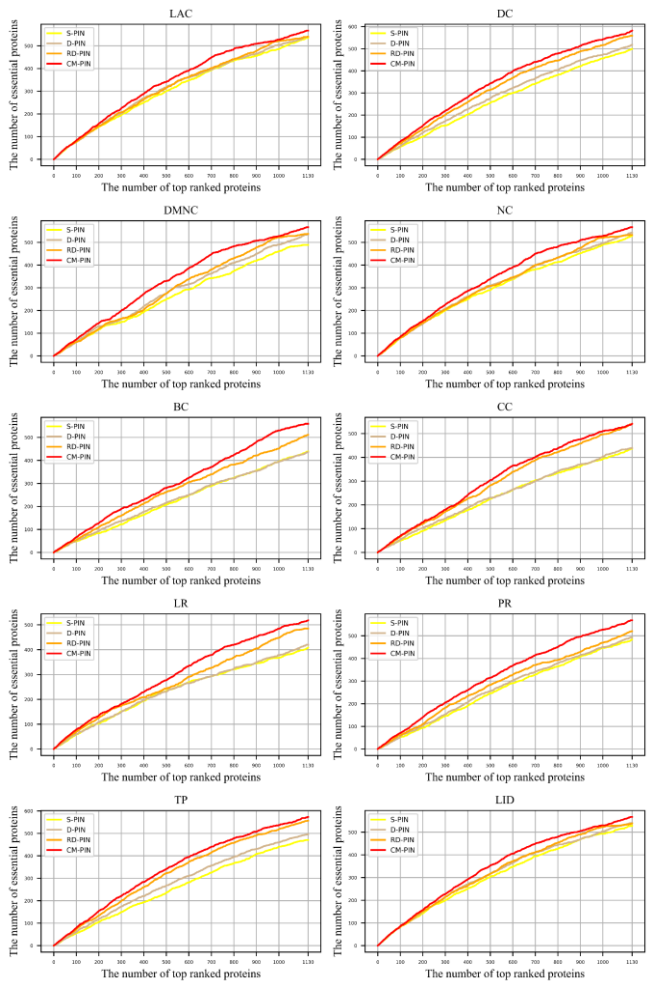

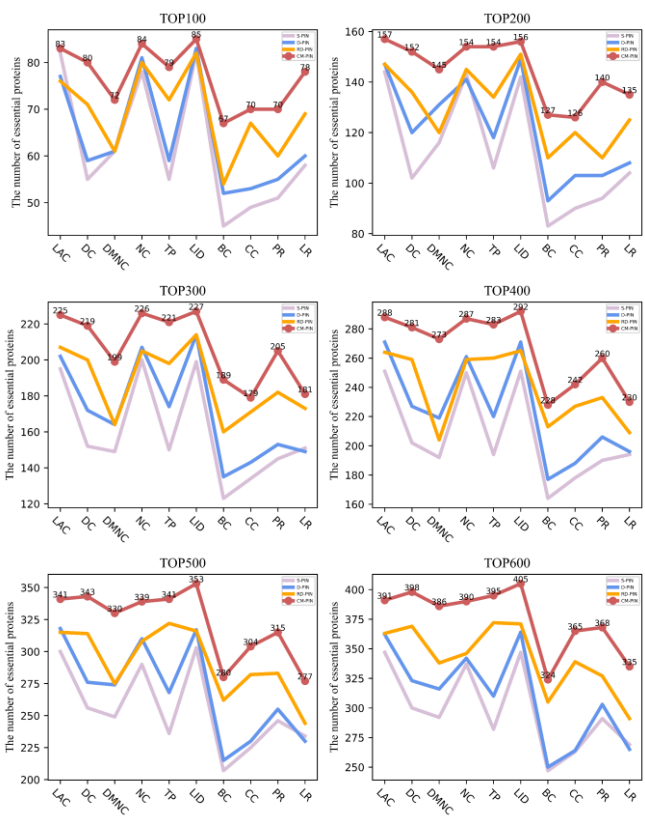

Fig.2. Comparison of the number of essential proteins identified by centrality methods in S-PIN, D-PIN, RD-PIN and CM-PIN on DIP dataset.

### 3.3.2 Validated by using the jackknifing methodology

In order to evaluate the overall performance of CM-PIN more comprehensively, we used the jackknifing method. Figure 3 showed the number of essential proteins in the top 1130 highest scoring proteins for each centrality method in S-PIN, D-PIN, RD-PIN and CM-PIN (1130 proteins were essential in the dataset DIP). The horizontal axis of the jackknifing plot indicates the number of proteins that ranked high in the network and the vertical axis represents the number of essential proteins among these top-ranked proteins. It is obvious that in CM-PIN, the jackknifing curves of the 10 centrality methods are all above the other three networks, and the differences are

Fig.3. Centrality methods are validated by the jackknife methodology on DIP dataset.

### 3.3.3 Analysis of precision-recall curves

As the identification of essential proteins is a sample imbalance problem, the number of negative class samples (non-essential proteins) is much larger than the number of positive class samples (essential proteins). Therefore, to assess the significance of CM-PIN, we used precision-recall curves to compare the efficiency of essential protein identification [44]. The vertical axis (precision) of the precision-recall curve reflects the proportion of the true positive examples in the positive examples determined by the classifier, and the horizontal axis (recall) reflects the proportion of the positive examples determined by the classifier in the total positive examples. What's more, we further calculated the area under the precision-recall curve (PRAUC), as shown in Figure 4 (the PRAUC values for each centrality method are shown in the legends), and it can be seen that both the precision-recall curves and

PRAUC values on the CM-PIN were the best. For the DIP dataset, the PRAUC values for the 10 centrality methods in the CM-PIN improved in a range of 4.7% to 15.8% compared to the RD-PIN which was the best of the other three networks.

### 3.3.4 Validated by accuracy

To further evaluate the overall performance of CM-PIN and the accuracy of essential protein identification, we used the following six evaluation metrics: sensitivity (SN), specificity (SP), positive predictive value (PPV), negative predictive value (NPV), F-measure (FM), and accuracy (ACC). The top 1130 proteins (1130 is the number of essential proteins for DIP) after the descending order of each centrality metric value were assumed to be essential proteins, and the calculation formulas are as follows, where TP is the correctly predicted essential protein, FP stands for the incorrectly predicted essential protein, TN refers to the correctly predicted non-essential protein, and FN represents the incorrectly predicted non-essential protein.

$$SN = \frac{TP}{TP + FN} \quad (15)$$

$$SP = \frac{TN}{FP + TN} \quad (16)$$

$$PPV = \frac{TP}{TP + FP} \quad (17)$$

$$NPV = \frac{TN}{TN + FN} \quad (18)$$

$$FM = \frac{2 \times SN \times PPV}{SN + PPV} \quad (19)$$

$$ACC = \frac{TP + TN}{TP + TN + FP + FN} \quad (20)$$

Table 4 showed the comparison results of the 10 centrality methods on the six indicators of S-PIN, D-PIN, RD-PIN and CM-PIN. It can be seen that the six evaluation indicators and top 1130 of the 10 centrality methods in CM-PIN are better than the other three networks, which indicates that the method of refining networks by modules in this paper is feasible and can effectively improve the identification accuracy of essential proteins.

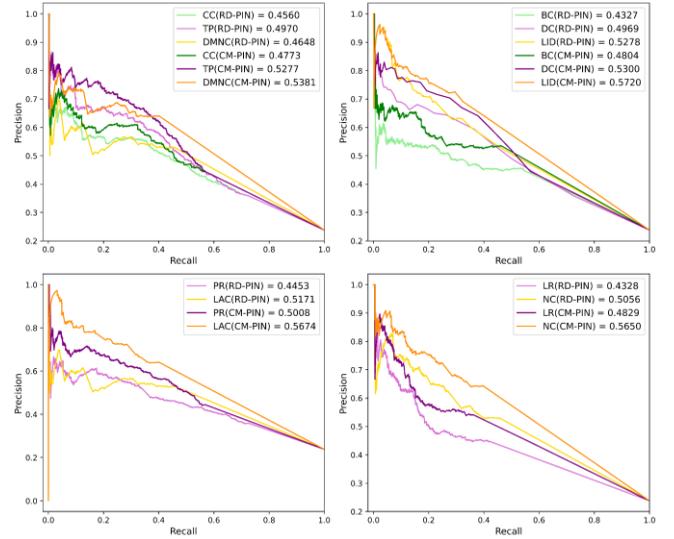

Fig.4. Comparison of precision-recall curves of different centrality methods in RD-PIN and CM-PIN on DIP dataset. The PRAUC values for each centrality method are shown in the legends.

Table 4. Comparison of evaluation indices for centrality methods on DIP dataset.

| Centrality | PIN | SN | SP | PPV | NPV | FM | ACC | Top 1130 |
|---|---|---|---|---|---|---|---|---|
| LAC | S-PIN | 0.4735 | 0.8355 | 0.4735 | 0.8355 | 0.4735 | 0.7493 | 535 |
| | D-PIN | 0.4779 | 0.8368 | 0.4779 | 0.8368 | 0.4779 | 0.7514 | 540 |
| | RD-PIN | 0.4779 | 0.8368 | 0.4779 | 0.8368 | 0.4779 | 0.7514 | 540 |
| | CM-PIN | **0.5018** | **0.8443** | **0.5018** | **0.8443** | **0.5018** | **0.7627** | **567** |
| DC | S-PIN | 0.4416 | 0.8255 | 0.4416 | 0.8255 | 0.4416 | 0.7341 | 499 |
| | D-PIN | 0.4584 | 0.8308 | 0.4584 | 0.8308 | 0.4584 | 0.7421 | 518 |
| | RD-PIN | 0.4947 | 0.8421 | 0.4947 | 0.8421 | 0.4947 | 0.7594 | 559 |
| | CM-PIN | **0.5150** | **0.8485** | **0.5150** | **0.8485** | **0.5150** | **0.7691** | **582** |
| DMNC | S-PIN | 0.4327 | 0.8227 | 0.4327 | 0.8227 | 0.4327 | 0.7299 | 489 |
| | D-PIN | 0.4735 | 0.8355 | 0.4735 | 0.8355 | 0.4735 | 0.7493 | 535 |
| | RD-PIN | 0.4761 | 0.8363 | 0.4761 | 0.8363 | 0.4761 | 0.7505 | 538 |
| | CM-PIN | **0.5018** | **0.8443** | **0.5018** | **0.8443** | **0.5018** | **0.7627** | **567** |
| NC | S-PIN | 0.4681 | 0.8338 | 0.4681 | 0.8338 | 0.4681 | 0.7467 | 529 |
| | D-PIN | 0.4805 | 0.8377 | 0.4805 | 0.8377 | 0.4805 | 0.7526 | 543 |
| | RD-PIN | 0.4717 | 0.8349 | 0.4717 | 0.8349 | 0.4717 | 0.7484 | 533 |
| | CM-PIN | **0.5018** | **0.8443** | **0.5018** | **0.8443** | **0.5018** | **0.7627** | **567** |
| TP | S-PIN | 0.4159 | 0.8175 | 0.4159 | 0.8175 | 0.4159 | 0.7219 | 470 |
| | D-PIN | 0.4389 | 0.8247 | 0.4389 | 0.8247 | 0.4389 | 0.7328 | 496 |
| | RD-PIN | 0.4920 | 0.8413 | 0.4920 | 0.8413 | 0.4920 | 0.7581 | 556 |
| | CM-PIN | **0.5071** | **0.846** | **0.5071** | **0.846** | **0.5071** | **0.7653** | **573** |
| LID | S-PIN | 0.4673 | 0.8335 | 0.4673 | 0.8335 | 0.4673 | 0.7463 | 528 |
| | D-PIN | 0.4779 | 0.8368 | 0.4779 | 0.8368 | 0.4779 | 0.7514 | 540 |
| | RD-PIN | 0.4761 | 0.8363 | 0.4761 | 0.8363 | 0.4761 | 0.7505 | 538 |
| | CM-PIN | **0.5018** | **0.8443** | **0.5018** | **0.8443** | **0.5018** | **0.7627** | **567** |
| BC | S-PIN | 0.3885 | 0.8089 | 0.3885 | 0.8089 | 0.3885 | 0.7088 | 439 |
| | D-PIN | 0.3858 | 0.8081 | 0.3858 | 0.8081 | 0.3858 | 0.7075 | 436 |
| | RD-PIN | 0.4522 | 0.8288 | 0.4522 | 0.8288 | 0.4522 | 0.7391 | 511 |
| | CM-PIN | **0.4947** | **0.8421** | **0.4947** | **0.8421** | **0.4947** | **0.7594** | **559** |
| CC | S-PIN | 0.3858 | 0.8081 | 0.3858 | 0.8081 | 0.3858 | 0.7075 | 436 |
| | D-PIN | 0.3885 | 0.8089 | 0.3885 | 0.8089 | 0.3885 | 0.7088 | 439 |
| | RD-PIN | 0.4779 | 0.8368 | 0.4779 | 0.8368 | 0.4779 | 0.7514 | 540 |
| | CM-PIN | **0.4788** | **0.8371** | **0.4788** | **0.8371** | **0.4788** | **0.7518** | **541** |
| LR | S-PIN | 0.3611 | 0.8003 | 0.3611 | 0.8003 | 0.3611 | 0.6957 | 408 |
| | D-PIN | 0.3717 | 0.8037 | 0.3717 | 0.8037 | 0.3717 | 0.7008 | 420 |
| | RD-PIN | 0.4292 | 0.8216 | 0.4292 | 0.8216 | 0.4292 | 0.7282 | 485 |
| | CM-PIN | **0.4584** | **0.8308** | **0.4584** | **0.8308** | **0.4584** | **0.7421** | **518** |
| PR | S-PIN | 0.4274 | 0.8211 | 0.4274 | 0.8211 | 0.4274 | 0.7273 | 483 |
| | D-PIN | 0.4363 | 0.8238 | 0.4363 | 0.8238 | 0.4363 | 0.7316 | 493 |
| | RD-PIN | 0.4602 | 0.8313 | 0.4602 | 0.8313 | 0.4602 | 0.7429 | 520 |
| | CM-PIN | **0.5027** | **0.8446** | **0.5027** | **0.8446** | **0.5027** | **0.7632** | **568** |

## 3.4 Selection and analysis of thresholds

In the DIP dataset, the optimal partitioning of modules was achieved by the Fast-unfolding algorithm when the modularity $Q = 0.7408$, at which point the RD-PIN was partitioned into 26 modules. We calculated three metrics for each module in RD-PIN: $Corr$, $NSL$, and $TF$ (as shown in Table 5) by using the biological information of the proteins and the topological information of the modules in the network. We also observed the number and proportion of essential proteins in each module and found that there was variation between modules and that some modules with sparse interactions within modules or with little biologically important information contained few essential proteins, which may be the non-critical modules or contain a lot of noise in modules. For example, modules 1, 24, and 26 have an $NSL$ of 0, which means that the proteins in their modules do not appear in the subcellular compartments of the nucleus, and after the thresholds screening, they will likely be defined as non-critical modules. Therefore, in order to get a more effective network, we need to identify the non-critical modules and to remove the interactions in these non-critical modules.

To obtain the variation rule of the effect of thresholds on the selection of critical modules and the performance of the network, according to the data distribution of three metrics in the module, we let $th_1 \in \{-0.02, -0.005, 0.015\}$, $th_2 \in \{1.5, 2\}$, $th_3 \in \{0.25, 0.5\}$, and listed the effect of the networks on the identification accuracy of essential proteins with different values of the thresholds, respectively (as shown in Table 6, the experimental results in the table are the performance of LID in different networks). The experimental results showed that when $th_1$ and $th_2$ were small and $th_3$ was large, more critical modules were selected. At this time, there was still a large amount of noise in the network that had not been eliminated and the improvement in identification accuracy of essential proteins was not significant, for example, when $th_1 = -0.02$, $th_2 = 1.5$ and $th_3 = 0.5$, the identification accuracy of essential proteins at top 600 and PRAUC have improved compared with RD-PIN, but the identification accuracy of essential proteins at top 1130 is not as good as RD-PIN. In contrast, when $th_1$ and $th_2$ were larger, fewer critical modules were selected. At this time, critical parts of the network may have been removed, and the improvement in the network's identification accuracy of essential proteins was not optimal, for example, when $th_1 = 0.015$, $th_2 = 2$ and $th_3 = 0.5$, the identification accuracy of essential proteins at top 1130 of LID in CM-PIN was still inferior to RD-PIN. Among them, the change of $th_1$ and $th_2$ has a greater impact on the selection of modules, because biological information can better assist in identifying essential proteins than the topology information of the network. When $th_1 = -0.005$, $th_2 = 2$ and $th_3 = 0.25$, the optimal CM-PIN in the DIP dataset is obtained.

In addition, in order to discuss the reason why the identification accuracy of essential proteins of each centrality method in CM-PIN is higher than that in the other three networks (S-PIN, D-PIN, RD-PIN), we also calculated the ratio of essential proteins in different proteins at top 1130 of each centrality method in CM-PIN and other three networks, as shown in Figure 5. It can be seen that in CM-PIN, each centrality method can identify some different essential proteins that cannot be identified in the other three networks. Even compared with the best RD-PIN in the three networks, some centrality methods can identify a large part of different essential proteins at top 1130 in CM-PIN, such as BC, which can identify 16% of the different essential proteins in CM-PIN that cannot be identified in RD-PIN. Therefore, the essential protein identification accuracy in CM-PIN is optimal for each centrality method.

Table 5. Biological and topological characterization of each module in RD-PIN on DIP dataset.

| Modules | $Corr$ | $NSL$ | $TF$ | Number of proteins/essential proteins |
|---|---|---|---|---|
| 1 | -0.0258 | 0 | 1.6667 | 33/3 |
| 2 | -0.0764 | 2.1847 | 0.4775 | 222/78 |
| 3 | -0.0075 | 0.1563 | 0.875 | 32/5 |
| 4 | -0.023 | 0.925 | 1.125 | 40/7 |
| 5 | 0.1688 | 3.12 | 2.68 | 175/128 |
| 6 | -0.0314 | 1.9419 | 0.2674 | 86/28 |
| 7 | -0.016 | 1.4701 | 0.3846 | 117/27 |
| 8 | 0.0362 | 2.2596 | 0.5096 | 104/40 |
| 9 | -0.0684 | 2.5321 | 0.9423 | 156/50 |
| 10 | -0.0358 | 0.4701 | 1.3806 | 134/33 |
| 11 | 0.1013 | 0.0467 | 1.6822 | 107/42 |
| 12 | 0.0317 | 2.8403 | 1.6736 | 144/76 |
| 13 | -0.001 | 2.2963 | 0.1481 | 27/9 |
| 14 | 0.0824 | 2.01 | 1.33 | 100/43 |
| 15 | -0.0314 | 2.4423 | 0.8077 | 52/18 |
| 16 | 0.0074 | 0.1111 | 2.5556 | 9/6 |
| 17 | -0.0493 | 2.3765 | 0.6 | 85/39 |
| 18 | 0.0612 | 2.8462 | 1.2051 | 78/38 |
| 19 | 0.0609 | 0.0962 | 1.1731 | 52/15 |
| 20 | -0.0736 | 2.8298 | 0.2128 | 47/14 |
| 21 | -0.0989 | 2.0381 | 0.8095 | 105/45 |
| 22 | 0.0264 | 1.9286 | 1 | 28/15 |
| 23 | -0.0208 | 1.6 | 0.4 | 20/1 |
| 24 | -0.0088 | 0 | 1.3929 | 28/2 |
| 25 | -0.0543 | 0.0857 | 1.8 | 35/1 |
| 26 | -0.0459 | 0 | 0.8333 | 6/0 |

### 3.5 Validated by the BioGRID dataset

To demonstrate that the network refinement approach of this paper is also valid on other datasets, we further performed related experiments on a larger protein interaction network, the BioGRID dataset [45], which contains 5,616 proteins, 52,833 pairs of protein interactions and contains 1,199 essential proteins. In BioGRID, 19 different modules were delineated by the Fast-unfolding algorithm, and the modularity ($Q$) is 0.6532 at this time. By analyzing the biological and topological characteristics in each module, we selected 15 critical modules and constructed the CM-PIN. Among them, the thresholds selection was as follows: $th_1$ = -0.015, $th_2$ = 1, $th_3$ = -2.

As shown in Table 7, we compared the PRAUC values, accuracy, and the identification accuracy of essential proteins at top 100, top 600, and top 1199 of each centrality method in CM-PIN and RD-PIN (the best of the other three networks). The results showed that all centrality methods had higher PRAUC values and accuracy in the higher quality network CM-PIN, and the identification accuracy of essential proteins at top 100, top 600, and top 1199 of each centrality is preferable. Therefore, these are further demonstrated that the network refinement method proposed in this paper can effectively remove false positives and noise from the already refined network and improve the identification accuracy of essential proteins.

### 3.6 Applying network refinement methods to S-PIN and D-PIN

This paper proposed a network refinement method and further constructs a more refined and effective network CM-PIN on RD-PIN, which demonstrated that the proposed network refinement method can further filter the noise in the network on the high-quality network and thus improve the identification accuracy of the essential proteins. However, to demonstrate whether the network refinement method can be effectively applied to networks with more noise and false positives, even PINs obtained by high-throughput biological experiments, we applied the network refinement method proposed in this paper to S-PIN and D-PIN of two datasets and compared the identification accuracies of essential proteins by 10 centrality methods, as shown in Table 8 and Table 9.

It can be seen that the ACC and PRAUC of the 10 centralities on the refined network constructed by the network refinement method in this paper are the highest, and the improvement effect is very significant. For example, in BioGRID dataset, compared with S-PIN and D-PIN, the number of essential proteins identified by CC method at top 1130 in each refinement network can be improved by 214 and 143 respectively, and its PRAUC can be improved by 54.8% and 38.4%, as we know, at present, there is no network refinement method that can improve the identification accuracy of the centralities for essential proteins such greatly. This undoubtedly proves that the network refinement method proposed in this paper can be effectively used in the network with a lot of noise and can improve the identification accuracy of the centralities for essential proteins.

Table 6. The variation of the effect of thresholds on the selection of critical modules and the performance of the network.

| $th_1$ | $th_2$ | $th_3$ | Number of critical modules | Top 100, 600,1130 | ACC | PRAUC |
|---|---|---|---|---|---|---|
| 0.015 | 1.5 | 0.25 | 15 | 84, 398, 563 | 0.7611 | 0.5636 |
| 0.015 | 2 | 0.25 | 13 | 84, 398, 551 | 0.7560 | 0.5685 |
| 0.015 | 1.5 | 0.5 | 13 | 85, 387, 532 | 0.7480 | 0.5548 |
| 0.015 | 2 | 0.5 | 15 | 83, 383, 533 | 0.7484 | 0.5498 |
| -0.005 | 1.5 | 0.25 | 17 | 85, 396, 567 | 0.7627 | 0.5672 |
| **-0.005** | **2** | **0.25** | **15** | **85, 405, 567** | **0.7627** | **0.5720** |
| -0.005 | 1.5 | 0.5 | 15 | 84, 395, 541 | 0.7518 | 0.5585 |
| -0.005 | 2 | 0.5 | 17 | 86, 387, 541 | 0. 7518 | 0.5536 |
| -0.02 | 1.5 | 0.25 | 20 | 84, 384, 556 | 0.7581 | 0.5539 |
| -0.02 | 2 | 0.25 | 18 | 85, 387, 555 | 0.7577 | 0.5578 |
| -0.02 | 1.5 | 0.5 | 17 | 85, 382, 531 | 0.7476 | 0.5512 |
| -0.02 | 2 | 0.5 | 19 | 86, 383, 535 | 0.7493 | 0.5467 |

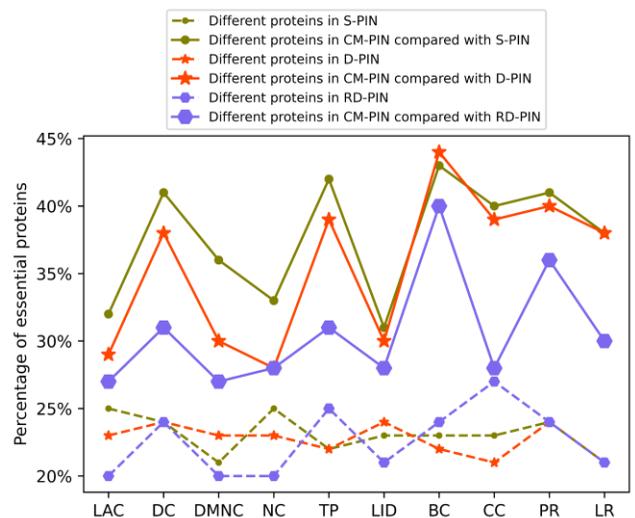

Fig.5. The comparison of the percentage of essential proteins in CM-PIN with that in other three networks in different proteins for each centrality method on DIP dataset.

Table 7. Comparison of PRAUC values of different centrality methods in RD-PIN and CM-PIN on BioGRID dataset.

| Centralities | RD-PIN | | | | | CM-PIN | | | | |
|---|---|---|---|---|---|---|---|---|---|---|
| | PRAUC | ACC | Top 100 | Top 600 | Top 1199 | PRAUC | ACC | Top 100 | Top 600 | Top 1199 |
| LAC | 0.4483 | 0.7824 | 57 | 339 | 588 | **0.4887** | **0.7927** | 57 | **380** | **617** |
| DC | 0.4744 | 0.7853 | 61 | 341 | 596 | **0.4938** | **0.7949** | 62 | **367** | **623** |
| DMNC | 0.4043 | 0.7653 | 28 | 307 | 540 | **0.4513** | **0.7756** | 55 | **350** | **569** |
| NC | 0.4607 | 0.7824 | 56 | 347 | 588 | **0.4990** | **0.7920** | 57 | **379** | **615** |
| TP | 0.4626 | 0.7642 | **70** | 339 | 537 | **0.4694** | **0.7721** | 69 | **341** | **559** |
| BC | 0.3789 | 0.7475 | 41 | 260 | 490 | **0.3990** | **0.7550** | 47 | **279** | **511** |
| CC | 0.3980 | 0.7482 | 50 | 273 | 492 | **0.4002** | **0.7514** | 50 | 273 | **501** |
| PR | 0.4419 | 0.7689 | 61 | 307 | 550 | **0.4677** | **0.7845** | 64 | **336** | **594** |
| LR | 0.3735 | 0.7372 | 56 | 253 | 461 | **0.4012** | **0.7450** | 58 | **266** | **483** |
| LID | 0.4509 | 0.7867 | 57 | 340 | 600 | **0.4912** | **0.7934** | 57 | **374** | **619** |

Table 8. Applying network refinement methods to S-PIN and D-PIN on DIP dataset.

| Centralities | S-PIN | | Refined network in S-PIN | | D-PIN | | Refined network in D-PIN | |
|---|---|---|---|---|---|---|---|---|
| | ACC (Top 1130) | PRAUC | ACC (Top 1130) | PRAUC | ACC (Top 1130) | PRAUC | ACC (Top 1130) | PRAUC |
| LAC | 0.7493 (535) | 0.4837 | **0.7535 (545)** | **0.5020** | 0.7514 (540) | 0.5030 | **0.7535 (545)** | **0.5371** |
| DC | 0.7341 (499) | 0.4111 | **0.7497 (536)** | **0.4589** | 0.7421 (518) | 0.4354 | **0.7556 (550)** | **0.4856** |
| DMNC | 0.7299 (489) | 0.4209 | **0.7509 (539)** | **0.4540** | 0.7493 (535) | 0.4512 | **0.7539 (546)** | **0.4999** |
| NC | 0.7467 (529) | 0.4775 | **0.7543 (547)** | **0.5055** | 0.7526 (543) | 0.4963 | **0.7531 (544)** | **0.5348** |
| TP | 0.7219 (470) | 0.3872 | **0.7488 (534)** | **0.4480** | 0.7328 (496) | 0.4194 | **0.7480 (532)** | **0.4782** |
| BC | 0.7088 (439) | 0.3498 | **0.7261 (480)** | **0.4136** | 0.7075 (436) | 0.3629 | **0.7269 (482)** | **0.4259** |
| CC | 0.7075 (436) | 0.3596 | **0.7354 (502)** | **0.4107** | 0.7088 (439) | 0.3719 | **0.7328 (496)** | **0.4278** |
| PR | 0.7273 (483) | 0.3879 | **0.7442 (523)** | **0.4432** | 0.7316 (493) | 0.3997 | **0.7429 (520)** | **0.4528** |
| LR | 0.6957 (408) | 0.3616 | **0.7269 (482)** | **0.4231** | 0.7008 (420) | 0.3733 | **0.7311 (492)** | **0.4264** |
| LID | 0.7463 (528) | 0.4881 | **0.7573 (554)** | **0.5042** | 0.7514 (540) | 0.5132 | **0.7539 (546)** | **0.5430** |

Table 9. Applying network refinement methods to S-PIN and D-PIN on BioGRID dataset.

| Centralities | S-PIN | | Refined network in S-PIN | | D-PIN | | Refined network in D-PIN | |
|---|---|---|---|---|---|---|---|---|
| | ACC (Top 1199) | PRAUC | ACC (Top 1199) | PRAUC | ACC (Top 1199) | PRAUC | ACC (Top 1199) | PRAUC |
| LAC | 0.7710 (556) | 0.4132 | **0.7799 (581)** | **0.4463** | 0.7756 (569) | 0.4311 | **0.7853 (596)** | **0.4692** |
| DC | 0.7568 (516) | 0.4096 | **0.7867 (600)** | **0.4444** | 0.7671 (545) | 0.4256 | **0.7821 (587)** | **0.4554** |
| DMNC | 0.7208 (415) | 0.3285 | **0.7660 (542)** | **0.4125** | 0.7461 (486) | 0.3760 | **0.7671 (545)** | **0.4362** |
| NC | 0.7724 (560) | 0.4235 | **0.7810 (584)** | **0.4590** | 0.7749 (567) | 0.4421 | **0.7856 (597)** | **0.4797** |
| TP | 0.7212 (416) | 0.3210 | **0.7742 (565)** | **0.4288** | 0.7240 (424) | 0.3384 | **0.7710 (556)** | **0.4304** |
| BC | 0.7407 (471) | 0.3636 | **0.7728 (561)** | **0.4416** | 0.7318 (446) | 0.3536 | **0.7575 (518)** | **0.4093** |
| CC | 0.6841 (312) | 0.2599 | **0.7603 (526)** | **0.4022** | 0.6952 (343) | 0.2752 | **0.7461 (486)** | **0.3809** |
| PR | 0.7589 (522) | 0.4035 | **0.7849 (595)** | **0.4520** | 0.7610 (528) | 0.4081 | **0.7853 (596)** | **0.4450** |
| LR | 0.7076 (378) | 0.3124 | **0.7468 (488)** | **0.3886** | 0.7112 (388) | 0.3271 | **0.7404 (470)** | **0.3755** |
| LID | 0.7721 (559) | 0.4182 | **0.7821 (587)** | **0.4479** | 0.7746 (566) | 0.4350 | **0.7867 (600)** | **0.4716** |

## 4 CONCLUSION

In this paper, we proposed a protein interaction network refinement method based on modular discovery and biological information. By dividing the network into different modules through module discovery algorithm and analyzing the homology information and the subcellular localization information of the proteins in each module, and the topological information of the modules in the network, we can construct a more detailed and effective network CM-PIN. In order to validate the effectiveness of this method, first, we evaluated and analyzed the network performance in DIP dataset with 10 typical network-based centrality methods (LAC, DC, DMNC, NC, TP, LID, CC, BC, PR, LR), and then verified it in BioGRID dataset, another larger dataset. The experimental results showed that the network refinement method in this paper can effectively filter the false positives and noise in the networks whether they are networks with a lot of noise or refined networks, and the constructed CM-PIN can effectively improve the identification accuracy of essential proteins for centrality method. The identification performances of the centrality methods in CM-PIN were better than that of the static network S-PIN and the two current-ly proposed networks refined with biological information: D-PIN and RD-PIN.

## ACKNOWLEDGMENT

Li Pan and Haoyue Wang contributed equally to this work.

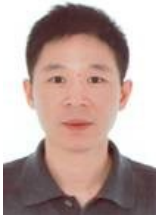

**LI PAN** received the Ph.D. degree in computer science from Tongji University, China, in 2009. He is currently a Professor with the School of Information Science and Engineering, Hunan Institute of Science and Technology, Yueyang, Hunan, China. His research interests include bioinformatics and Petri nets.

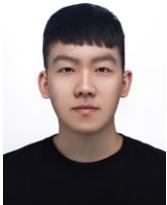

**HAOYUE WANG** is currently pursuing the Graduate degree with the Hunan Institute of Science and Technology. His research interests include bioinformatics and complex networks.

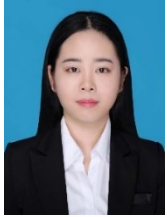

**JING SUN** is currently pursuing the Graduate degree with the Hunan Institute of Science and Technology. Her research interests include bioinformatics and complex networks.

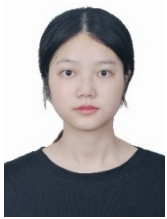

**BIN LI** is currently pursuing the Graduate degree with the Hunan Institute of Science and Technology. Her research interests include bioinformatics and complex networks.

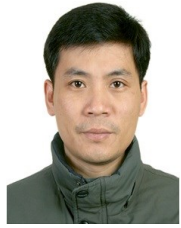

**BO YANG** received the B.Sc. degree in mechanical engineering from Zhengzhou University, China, in 1996, the M.Sc. degree in computer application technology from Xiangtan University, China, in 2004, and the Ph.D. degree in mechanical and electrical engineering from Central South University, China, in 2010. Since 2012, he has been an Associate Professor with the College of Information Science and Technology, Hunan Institute of Science and Technology. His research interests include MR brain image analysis, statistical pattern recognition, and machine learning

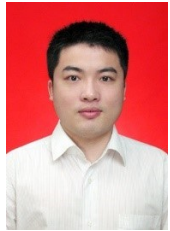

**WENBIN LI** received the B.S. degree in computer science and technology from Hunan Normal University, Changsha, China, in 2003, the M.S. degree in computer applications technology from the Changsha University of Science and Technology, Changsha, in 2006, and the Ph.D. degree in control engineering from Central South University, Changsha, in 2020. His research interests include industrial process control, evolutionary computation, and multi-objective optimization.